# Self-Heterodyned Detection of Dressed State Coherences in Helium by Noncollinear Extreme Ultraviolet Wave Mixing with Attosecond Pulses

**Ashley P. Fidler[1,2], Erika R. Warrick[1,2], Hugo J. B. Marroux[1,2], Etienne Bloch[1], Daniel M. Neumark[1,2], and Stephen R. Leone[1,2,3]**

[1] Chemical Sciences Division, Lawrence Berkeley National Laboratory, Berkeley, CA 94720, USA

[2] Department of Chemistry, University of California, Berkeley, CA 94720, USA

[3] Department of Physics, University of California, Berkeley, CA 94720, USA

E-mail: srl@berkeley.edu

## Abstract

Noncollinear wave-mixing spectroscopies with attosecond extreme ultraviolet (XUV) pulses provide unprecedented insight into electronic dynamics. In infrared and visible regimes, heterodyne detection techniques utilize a reference field to amplify wave-mixing signals while simultaneously allowing for phase-sensitive measurements. Here, we implement a self-heterodyned detection scheme in noncollinear wave-mixing measurements with a short attosecond XUV pulse train and two few-cycle near infrared (NIR) pulses. The initial spatiotemporally overlapped XUV and NIR pulses generate a coherence of both odd (*1snp*) and even (*1sns* and *1snd*) parity states within gaseous helium. A variably delayed noncollinear NIR pulse generates angularly-dependent four-wave mixing signals that report on the evolution of this coherence. The diffuse angular structure of the XUV harmonics underlying these emission signals is used as a reference field for heterodyne detection, leading to cycle oscillations in the transient wave-mixing spectra. With this detection scheme, wave-mixing signals emitting from at least eight distinct light-induced, or dressed, states can be observed, in contrast to only one light induced state identified in a similar homodyne wave-mixing measurement. In conjunction with the self-heterodyned detection scheme, the noncollinear geometry permits the conclusive identification and angular separation of distinct wave-mixing pathways, reducing the complexity of transient spectra. These results demonstrate that the application of heterodyne detection schemes can provide signal amplification and phase-sensitivity, while maintaining the versatility and selectivity of noncollinear attosecond XUV wave-mixing spectroscopies. These techniques will be important tools in the study of ultrafast dynamics within complex chemical systems in the XUV regime.

## 1. Introduction

Spectroscopy in the extreme ultraviolet (XUV) and x-ray regimes provides fundamental insights into the nature of matter through its element specificity and charge-state sensitivity. Sub-optical wavelength light sources access high-energy electronic transitions, including inner valence and core-level excitations that decay via electronic correlation on timescales of hundreds of attoseconds to a few femtoseconds [1,2]. To probe electronic dynamics, attosecond pulses in the XUV and x-ray regimes have been produced with high harmonic generation (HHG)-based table-top set-ups [3,4] and free electron lasers [5–7]. Attosecond pulses have been successfully utilized to study chemically-signficant ultrafast dynamics, including bond cleavage following an avoided crossing in IBr [8], electronic structure changes in the aromatic biomolecule phenylalanine [9], and carrier-carrier interactions after excitation in silicon [10].

Nonlinear wave-mixing techniques are regularly employed at infrared and visible wavelengths to disentangle complex spectra in multiple dimensions, yielding time-resolved information about coupling in complicated ensembles [11–14]. These techniques can be broadly classified by their detection scheme. In the ultrafast community, homodyne detection typically refers to an experimental geometry in which the nonlinear wave-mixing signal is generated in the absence of another interfering laser pulse [12]. Using a square law detector, the intensity of the generated four-wave mixing signal is given by the following expression:

$$I_{signal} = |E_{emit}|^2 \tag{1}$$

where $I_{signal}$ is the intensity of the detected signal and $E_{emit}$ is the electric field of the four-wave mixing signal. In heterodyne detection, another pulse, often referred to as a local oscillator or reference field, interferes with the generated signal. The intensity of the signal detected in a heterodyne geometry can be described by the following expression:

$$I_{signal} = \left|E_{ref} + E_{emit}\right|^2 = \left|E_{ref}\right|^2 + |E_{emit}|^2 + 2\left|E_{ref}E_{emit}\right|\cos\left(\phi_{ref} - \phi_{emit}\right) \quad (2)$$

where $E_{ref}$ is the electric field of the reference pulse and $\phi_{ref}$ and $\phi_{emit}$ are the phases of the reference and emitted fields, respectively [15,16]. The reference field can either be independently supplied or, in a special case known as self-heterodyne detection, may consist of one of the pulses initially used to generate the nonlinear signal [12]. Although often more difficult to implement, heterodyne techniques amplify weak wave-mixing signals and allow for phase-sensitive detection [15–19].

Despite considerable theoretical effort [20,21], applications of nonlinear spectroscopic techniques in attosecond science have been limited by the dearth of reliable, high photon flux light sources and XUV/x-ray optics [16,22]. Nevertheless, attosecond transient absorption (ATA) with collinear XUV pump and optical or near infrared (NIR) probe pulses can be described as a self-heterodyned nonlinear spectroscopy [12]. These measurements provide evidence of four-wave mixing interactions in the form of quantum beat oscillations [22–26], but they lack the selectivity of higher-order wave-mixing techniques to disentangle complex spectra [27]. Capitalizing on the phase-matching conditions inherent in wave-mixing processes, recent experiments employed a noncollinear beam geometry between an XUV attosecond pulse train and either one [28,29] or two [30–32] moderately intense NIR pulses to measure transient spectra of spatially isolated XUV wave-mixing signals. These time-dependent wave-mixing signals were utilized to characterize a dark state in $N_2$ [30], to measure ultrafast lifetimes of autoionizing states in krypton [32], and to construct the first multidimensional spectrum in the XUV by implementing NIR pulse shaping techniques [33]. While highly versatile, these noncollinear wave-mixing measurements are typically best described as homodyne techniques [32], and therefore do not benefit from the signal enhancement and phase-sensitivity characteristic of heterodyne techniques.

Here, we implement an XUV wave-mixing spectroscopy with a variably delayed, noncollinear NIR pulse in a self-heterodyned detection scheme to differentiate distinct NIR-induced coupling pathways in the *1snl* Rydberg series of helium atoms. Time-dependent wave-mixing emission signals encoding the evolution a coherence of odd and even parity states initiated by a short XUV pulse train and a collinear, time-coincident NIR pulse arise from one-photon dipole-allowed *1snp* bright states and dressed states that appear in strong NIR fields, called light-induced states (LISs). Taking advantage of the angularly diffuse structure of the incident harmonics allows the wave-mixing signals to be superimposed onto a broadband XUV continuum, which acts as the reference field in a self-heterodyne detection scheme. Pronounced cycle oscillations, made possible by the underlying XUV reference field, reveal wave-mixing signals emitting from LISs that were not detected in previous measurements [31]. Moreover, the beam geometry and pulse sequence realize angle-dependent emission of distinct types of NIR-induced coupling pathways regardless of emission energy, demonstrating the high selectivity of these wave-mixing techniques. These results demonstrate that combining noncollinear wave-mixing geometries with heterodyne detection schemes confers tangible benefits in both sensitivity and selectivity, illuminating a mechanism by which weak XUV wave-mixing signals that encode ultrafast processes can be distinguished in complex dynamical systems.

## 2. Methods

The experimental apparatus employed for XUV wave-mixing measurements has been described previously and is shown in Fig. 1a [30–32]. Briefly, 22 fs duration NIR pulses produced by a 2 mJ, 1 kHz Ti:Sapphire multipass laser system (Femtopower HE, Femtolasers) are spectrally broadened in a homebuilt hollow core fiber statically filled with 1 bar of neon gas and then temporally compressed with six pairs of double angle chirped mirrors (PC70, Ultrafast Innovations). The resulting 6 fs broadband pulses (550 – 950 nm) are transmitted through a 50:50 beamsplitter and focused by a silver mirror (f = 50 cm) into a vacuum apparatus ($10^{-6}$ Torr) containing a 1-mm pathlength cell filled with xenon gas for HHG. Light passes through the glass cell via openings drilled on both sides of the barrel by the focused output of the amplifier. The few-cycle driving pulse generates a train of 2 – 3 subfemtosecond XUV pulses. Both the XUV and co-propogating NIR pulses are focused by a gold-coated toroidal mirror (f = 50 cm) into a 3-mm long target cell filled with 8 Torr of helium gas. The beam diameter of the collinear NIR pulse is approximately ~150 µm at full width half maximum (FWHM), which is substantially larger than the estimated ~20 µm XUV beam waist. Together, these co-propagating pulses are responsible for exciting states of both even and odd parity in the helium gas sample.

The NIR pulses reflected by the 50:50 beamsplitter allow for the measurement of time-domain spectra. These pulses are delayed relative to the co-propagating XUV / NIR pulses by a piezoelectric stage (P-622 with E509 controller, Physik Instrumente). A 1-m focal length mirror directs the NIR beam above the hole of an annular mirror in the XUV beampath before focusing it to a beam diameter of ~100 µm in the target cell. The noncollinear beam geometry is depicted in Fig. 1b. Temporal and spatial overlap between the HHG arm and the noncollinear NIR arm are determined by inserting a pick-off mirror after the annular mirror, sending the NIR pulse co-propagating with the XUV and the noncollinear NIR pulse to a CMOS camera (DCC1545M, Thorlabs) positioned outside of the vacuum apparatus at the focal position. The delay of the piezoelectric stage is varied until fringes in the spatial profile of the overlapped NIR beams appear, indicating temporal overlap has been obtained. A crossing angle of 1° (18 mrad) between the co-propagating XUV/NIR beams and the noncollinear NIR beam is estimated from the fringe spacing on the CMOS camera.

A 0.2 µm Al foil (Lebow) attenuates the NIR beams after the target cell. The transmitted XUV light is spectrally dispersed by an aberration-corrected concave grating (001-0464, Hitachi) and imaged onto a 1340 by 400 pixel X-ray charged-coupled device (CCD) camera (Pixis XO 400B, Princeton Instruments). The angular structure of the XUV harmonics is adjusted to interfere with generated wave-mixing signals on the detector by modifying the hollow core fiber mode, the position of the HHG cell relative to the focus of the NIR driver, and the xenon gas density used for HHG. Note that in this self-heterodyned geometry, the entire angular width of the XUV harmonics is perturbed by the helium gas, leading to absorption features even where the harmonics are diffuse.

The pulse sequence employed in these experiments is shown in Fig. 1c. A step size of 300 attoseconds was chosen for the delay between the time-coincident XUV/collinear NIR pulses and the noncollinear NIR pulse to resolve oscillations in the time-domain spectrum. Here, positive delays indicate that the co-propagating XUV and NIR pulses precede the noncollinear NIR pulse. A total of 1500 laser pulses were accumulated three times at each time delay to obtain a suitable signal-to-noise ratio. The horizontial energy axis of the X-ray CCD camera was calibrated daily with atomic transition line data available from the National Institute of Standards and Technology (NIST). Unless noted otherwise, all CCD camera images and time-domain spectra are plotted in absorbance (A)

$$\mathrm{A} = -\log_{10}\left(I/I_0\right) \tag{3}$$

where $I$ and $I_0$ represent the light transmitted through the sample cell with and without helium gas, respectively. For higher-order XUV wave-mixing features spatially isolated from the harmonics due to the noncollinear geometry, $I_0$ consists only of scattered light and dark counts on the camera. This presentation scheme was chosen to emphasize comparatively weak wave-mixing features and to differentiate absorption and emission signals.

## 3. Results

Wave-mixing signals are generated from the long-lived Rydberg states below the first ionization potential (IP) at 24.59 eV in helium gas. A short XUV pulse train generated by HHG and the time-coincident, collinear NIR driving pulse induce a coherent superposition of the ground state and both one-photon, dipole-allowed *1snp* (bright) and one-photon, dipole-forbidden *1sns* and *1snd* (dark) states, as demonstrated in previously reported four-wave mixing experiments [22,28,30]. A noncollinear few-cycle NIR pulse completes the wave-mixing scheme, producing angularly-dependent nonlinear signals that emit at the wavevector sum of the collinear excitation pulses ($k_{XUV}, k_{NIR\ 1}$) and noncollinear NIR ($k_{NIR\ 2}$) pulse.

A self-heterodyned detection scheme for the generated noncollinear wave-mixing signals is enabled by the angular structure of the harmonics. A CCD camera image of the spectrally dispersed 13$^{th}$ and 15$^{th}$ harmonics is shown in Fig. 2a. The camera image is plotted as a function of photon energy and divergence angle relative to the point of peak XUV intensity. Both harmonics exhibit significant angular dependence, tilting toward negative divergence angles with increasing photon energy. While the most intense region of these harmonics is relatively narrow (~1.4 mrad), a broad, diffuse background is observed in both the angular and energetic dimensions. In energy, the spectrally broad harmonic structure ensures continuum-like coherent excitation of the entire Rydberg series. In the angular dimension, this structure increases the effective span of the harmonics to over 4 mrad.

The addition of helium gas to the target cell results in multiple orders of wave-mixing signals at temporal overlap of all three pulses. A CCD camera image obtained with temporally overlapped XUV and NIR pulses is shown in Fig. 2b. Around 0 mrad, absorption features from the *1snp* bright states of helium can be observed in the spectrally dispersed harmonics. The moderately intense NIR pulses perturb the XUV-induced polarization that generates these features, leading to spectral broadening and AC-Stark shifting of the absorption lineshapes [26,34]. Cross section-dependent resonant pulse propogation effects further distort the absorption profile of the *1s2p* state [35–37]. Additional absorption features corresponding to LISs, which can be simplistically described as virtual states in a two-photon Raman-like transition to a dark state involving an XUV and a NIR photon, appear at energies distinct from the bright states [31,34,38–41]. As shown in Table 1, LISs appear in energy either one NIR photon above (+) or below (−) their corresponding dark states. Due to the electronic structure of helium, all but one of the *ns*+ or *nd*+ LISs arise around and above the IP. The *2s*+ and *ns*/*nd*– LISs can be observed at lower photon energies (19 – 23 eV).

As illustrated by Fig. 2c, the noncollinear beam geometry results in the emission of nonlinear wave-mixing signals at larger divergence angles. These wave-mixing signals are enhanced when each photon of the process corresponds to a resonant transition in the medium [42,43], leading to preferential emission at the energies of the *1snp* bright states and bright state-like LISs. At temporal overlap, multiple wave-mixing emission orders at the energies of the bright states are observed in the CCD camera image. Several of these emission orders are situated in regions with significant XUV flux due to the diffuse harmonic structure. Note that these emission features cannot be assigned to individual wave-mixing pathways, as wavevector phase-matching allows multiple, energetically-degenerate wave-mixing signals to emit at the same spatial location [32].

Four-wave mixing pathways that include one noncollinear NIR photon are emitting at divergence angles given by Eq. (4):

$$\phi_{div} \approx \frac{\upsilon_{NIR2}\theta_{NIR2}}{\upsilon_{XUV\pm NIR1}} \tag{4}$$

where $\upsilon_{NIR2}$ and $\theta_{NIR2}$ are the noncollinear NIR photon frequency and crossing angle, respectively, and $\upsilon_{XUV\pm NIR1}$ is the frequency associated with the two-photon excitation. Depending on the frequencies of the photons involved, four-wave mixing pathways of this type will typically emit ~1 – 1.7 mrad above and below the harmonic axis. Four-wave mixing pathways composed of two noncollinear NIR photons are also possible and emit at an angle described by a similar expression:

$$\phi_{div} \approx \frac{\upsilon_{NIR2}\theta_{NIR2} \pm \upsilon_{NIR2}\theta_{NIR2}}{\upsilon_{XUV}} \tag{5}$$

where $\upsilon_{XUV}$ is the frequency of the XUV excitation. Four-wave mixing pathways comprised of two noncollinear photons with the same wavevector emit ~2 – 3.2 mrad above and below the harmonic axis. Pathways comprised of noncollinear NIR photons with opposite wavevectors emit near the harmonic axis at 0 mrad. As the angular structure of the harmonics is imprinted on the wave-mixing signals, the effective width of the wave-mixing features extends beyond the emission ranges calculated above, ensuring that four-wave mixing signals at least partially overlap with the angularly diffuse harmonics.

Delaying the noncollinear NIR pulse relative to the two excitation pulses probes the evolution of the coherences. Because the collinear NIR photon is time-coincident with the XUV pulse, only wave-mixing processes completed by a noncollinear NIR photon produce delay-dependent wave-mixing signals. The temporal evolution of these signals can be assessed by integrating over a portion of their angular range and plotting the result as function of delay. In Fig. 3a-c, false color plots generated by integrating over the 0.65 mrad regions indicated in the CCD camera image (Fig. 2b) illustrate the key features of the angularly-dependent transient spectra. The spectrum taken from the region at the harmonic axis is dominated by absorption features (Fig. 3a). Unlike traditional ATA measurements, this experiment utilizes two distinct NIR pulses that arrive at different delays. The collinear NIR pulse ($\upsilon_{NIR1}$) co-propagating with the XUV pulse train perturbs the XUV-induced polarization of the bright states, broadening bright state absorption lineshapes, and generates LIS absorption features. These features are observed irrespective of delay because they arise exclusively from the time-coincident excitation (pump) pulses. The variably delayed, noncollinear NIR pulse ($\upsilon_{NIR2}$) is responsible for the time-dependent features in the spectrum, including additional broadening and AC-Stark shifting of the absorption features at overlap as well as oscillations at positive delays. Fast oscillations with a period of ~2.6 fs (1ω of the fundamental NIR light) can be distinguished throughout the spectrum at energies associated with both bright states and LISs. Slower modulations are readily observed in the *1s2p* absorption feature. As in previously reported collinear experiments [22,26], these oscillations are indicative of wave-mixing emission in a self-heterodyned detection scheme and will be described fully in the Discussion section.

Reduced XUV harmonic flux below (Fig. 3b) and above (Fig. 3c) the harmonic axis allows wave-mixing emission features to compete more effectively with absorption. These wave-mixing signals interfere with the residual XUV field, extending the self-heterodyned detection scheme to non-zero divergence angles. The noncollinear geometry not only permits strong wave-mixing emission to be superimposed on a weak XUV field relative to the on-axis harmonics, but also angularly differentiates wave-mixing pathways. In both off-axis transient spectra, emitted light from four-wave mixing processes can be oberved at *1snp* bright states and LIS energies. All wave-mixing features exhibit fast (~2.6 fs) modulations, similar to those observed on-axis. However, the intensity and slower (~27 fs) oscillatory structure of the wave-mixing signals varies with divergence angle. Above the harmonic axis (Fig. 3b), heavily-modulated wave-mixing signals preferentially emit from the *1s2p* state and nearby LISs. Only faint wave-mixing emission around time overlap can be identified for higher-lying states. Below the harmonic axis (Fig. 3c), intense long-lived wave-mixing emission with slow modulations can be observed from the *1snp* states and LISs near the IP. Although visible, wave-mixing near the *1s2p* state is less pronounced than in Fig. 3b.

Further insight into the long-lived four-wave mixing signals is obtained by taking the Fourier transform along the delay axis of each transient spectra, yielding the oscillation frequencies as a function of photon energy. The results are shown in Fig. 4a-c. For each of the divergence angle-dependent spectra, the fast 2.6 fs oscillations yield non-zero Fourier features situated at frequencies equivalent to one broadband NIR photon (~1.2 – 1.9 eV). At photon energies associated with the LISs, these non-zero Fourier features form lines of unity slope that, when traced to zero Fourier frequency, intercept the known energies of specific *1sns* and *1snd* (dark) states. Intercepted dark states include the *1s2s*, *1s3s*, *1s3d*, *1s4s*, *1s5s*, and *1s6s* states. While all Fourier plots exhibit unity slope lines, the sign of the slope and dark states intercepted differ between the three spectra. Both positive and negative unity slope lines at LIS energies can be identified at the harmonic axis (Fig 4a). Above the harmonic axis (Fig. 4b), the unity slope lines are exclusively negative, intercepting dark states situated above them in energy. These unity slope features are clustered in a spectral region associated with primarily *ns*– or *nd*– LISs just above the *1s2p* state in energy. Below the harmonic axis (Fig. 4c), unity slope lines are positive and are also observed around the IP, a spectral region corresponding to the *ns*+ or *nd*+ LISs.

Features around zero Fourier frequency contain signatures of the slower modulations. Along the harmonic axis (Fig. 4a), clusters of low frequency features are observed around and above the *1s2p* state (21.1 eV) as well as near the IP. These features exhibit a complex structure that includes unity slope lines extending to nearly 0.6 eV. Broad low-frequency Fourier features can be observed at photon energies with intense wave-mixing emission both above (Fig. 4b) and below axis (Fig. 4c). However, distinct peaks associated with the ~27 fs oscillations observed in the time-domain spectra are not well resolved due to the brevity of the measurement.

## 4. Discussion

Self-heterodyned noncollinear wave-mixing with attosecond XUV and few-cycle NIR pulses produces rich transient spectra dominated by fast and slow oscillations. The time-domain features depend not only on delay, but also on the divergence angle relative to the incident harmonics. The interpetation of these angularly-dependent spectra requires consideration of both the detection scheme and the noncollinear wave-mixing geometry.

### *4.1 Sensitive Self-Heterodyne Detection of Dressed State Coherences*

Pronounced oscillations in the transient wave-mixing spectra provide a window into the coherence initiated by two-photon excitation and NIR-induced coupling. Common to all observed wave-mixing features are fast (~2.6 fs) oscillations with frequencies equivalent to one NIR photon (1ω oscillations). Similar cycle oscillations have been reported in four-wave mixing measurements using analogous pulse sequences in a collinear beam geometry [22] and later at intermediate divergence angles in a noncollinear beam geometry [30]. Here, the cycle oscillatory features are directly attributed to the self-heterodyned detection scheme. The spatiotemporally overlapped XUV and NIR pulses generate a coherent superposition, or polarization, of dark and bright states. Wave-mixing emission occurs when a NIR photon couples one of these dark states to a bright state or LIS from which a transition to the ground state can occur. Several representative wave-mixing pathways are shown in Fig. 5. In a self-heterodyned scheme, wave-mixing emission from these bright states interferes with the XUV field transmitted through the sample, leading to the oscillatory stucture. The oscillation frequency corresponds to the energy difference between the dark state and the state from which the wave-mixing signal emits. As these states are coupled by a single NIR photon, cycle oscillations are observed.

Cycle quantum beating at LIS energies both along the harmonic axis and at larger divergence angles represents a particularly striking example of the interference effects in this self-heterodyned detection scheme. In the Fourier spectra, these fast oscillations produce intense unity slope features that span a frequency range greater than 0.5 eV and point to specific dark states when extended to zero Fourier frequency. The explanation for these features follows naturally from the characteristics of LISs. As LISs arise at an energy that is one NIR photon away from a dark state, their spectral width is determined by the bandwidth of the NIR pulse. The unity slope lines in the Fourier plot can therefore be ascribed to the coupling of a spectrally narrow dark state with a broad LIS by a broadband NIR pulse. The span of frequency values is dictated by both the collinear NIR pulse, as it establishes the spectral width of the LIS, and the noncollinear NIR pulse that couples the LIS with a dark state. The cycle oscillations persist past overlap because the wave-mixing signals encode the temporal evolution of the dark states, which do not decay appreciably on the timescale of the measurement [41].

Although difficult to differentiate in the time-domain plots, the Fourier spectra reveal wave-mixing emission from at least eight LISs. The dark states associated with these LIS are revealed explicitly by the intercept of the unity slope lines at zero Fourier frequency. Using this dark state information as well as the sign of the slope, wave-mixing emission from the *2s*+, *3s*+, *3s*–, *3d*+, *4s*+, *4s*–, *5s*+ and *6s*+ LISs can be identified conclusively. Notably, despite producing multiple orders of wave-mixing signal at LIS energies, a noncollinear experiment on the same system with time-coincident NIR pulses and angularly narrow harmonics obtained wave-mixing from only a single LIS [31]. Polarization measurements and theoretical calculations were required to identify this LIS definitively. In conjunction with the pulse sequence necessary for two-photon excitation, the self-heterodyne detection scheme therefore confers a two-fold advantage over homodyne wave-mixing techniques. First, the heterodyne scheme displays increased sensitivity to weak wave-mixing signals, such as emission from dressed states, compared to the homodyne case. Second, the interference oscillations indicative of the heterodyne scheme provide additional information about probe-induced coupling, which can be used to identify both intermediate states participating in the wave-mixing pathways and differentiate emitting states in complex, overlapping spectra.

Slow modulations in the transient spectra can also be ascribed to the interference of wave-mixing pathways. At the harmonic axis, a majority of low-frequency oscillations arise from the interference of wave-mixing pathways composed of two noncollinear NIR photons that emit on-axis (Fig. 2c) with XUV-generated absorption features. Above and below the harmonic axis, slow oscillations in spectral regions with intense wave-mixing emission are likely due to the interference of multiple wave-mixing pathways that incorporate one noncollinear NIR photon. As described in previous homodyne wave-mixing experiments [28,30], the interference of these emission pathways results in beating with frequencies corresponding to energy differences between dark states. While not well resolved in the Fourier spectra shown in Fig. 4, ~27 fs oscillations produce a Fourier peak at 0.15 eV, which corresponds to the energy difference between the *1s3d* and the *1s3s* dark states. These dark states can serve as intermediate states in wave-mixing pathways emitting from either the *1s2p* state or the Ryberg states near the IP.

### *4.2 Angle-Resolved Detection of Distinct Wave-Mixing Pathways*

The noncollinear beam geometry serves as an additional mechanism by which desired wave-mixing pathways can be differentiated in complex systems. Because the XUV and collinear NIR pulses arrive simultaneously, only wave-mixing pathways completed by the noncollinear NIR photon result in time-dependent emission. Moreover, the wavevector of the noncollinear NIR photon separates the emission from distinct time-dependent wave-mixing pathways. As demonstrated by the phase-matching diagrams in Fig. 5a, only wave-mixing pathways that include the absorption of a noncollinear NIR photon emit below the harmonic axis. For time-dependent wave-mixing signals, this constraint ensures that the intermediate dark state is located below the emitting bright state in energy. Alternatively, wave-mixing pathways that phase-match above the harmonic axis must emit a noncollinear photon (Fig. 5b). These distinctions are particularly apparent at energies associated with the LISs, where unity slope lines in the Fourier spectra dramatically change sign above and below the harmonic axis. Only emission from *ns*+ or *nd*+ LIS is observed below axis, while only *ns*– or *nd*– LIS can emit above axis. Due to the angular width of wave-mixing signals, both *ns*/*nd*+ and *ns*/*nd*– LISs can be distinguished at the harmonic axis.

The spatial separation of distinct types of wave-mixing pathways has been demonstrated in noncollinear XUV wave-mixing measurements with attosecond pulses [28,30], but not in a self-heterodyned or, more broadly, heterodyned geometry. Although technically heterodyned, collinear ATA and wave-mixing methods cannot utilize a spatial dimension to differentiate coupling pathways, often resulting in complex spectra. Note that the noncollinear geometry used in the experiments presented here effectively prohibits the interference of two wave-mixing pathways posited to result in cycle oscillations  in collinear measurements [22]. The angular separation of distinct types of wave-mixing pathways therefore supports the conclusion of the previous section, namely that the cycle modulations observed here are a consequence of the self-heterodyned geometry.

## 5. Conclusions

In summary, we demonstrate the exceptional sensitivity and selectivity of XUV wave-mixing spectroscopies with attosecond pulses in a self-heterodyned noncollinear geometry. Diffuse XUV harmonics serve as a reference field for divergence angle-dependent four-wave mixing emission signals generated in helium gas. These four-wave mixing signals encode the evolution of a bright (*1snp*) and dark (*1sns* and *1snd*) state coherence induced by spatiotemporally overlapped XUV and NIR pulses. A noncollinear few-cycle NIR pulse serves as the final interaction in the three-photon wave-mixing pathway, leading to time-dependent wave-mixing emission above and below the harmonic axis. Transient spectra taken of these wave-mixing signals reveal pronouced fast and slow modulations. While the slow oscillations arise from the interference of multiple wave-mixing pathways, the fast oscillations are ascribed to the interference of wave-mixing emission with the XUV reference field. The signal amplification due to this reference field allows for the observation and unambiguous identification of wave-mixing from at least eight different LIS states, in contrast to previous homodyned wave-mixing experiments on the same system [31].  Notably, these benefits in sensitivity are accrued without compromising the selectivity provided in noncollinear wave-mixing measurements. As shown here, a noncollinear beam geometry accesses an additional spatial dimension in which distinct wave-mixing pathways can be differentiated, reducing the number of interfering contributions that appear in an individual transient spectrum. In future measurements, diffraction of the harmonics before the sample can be used to produce an unperturbed XUV reference field, further decomplicating the obtained transient spectra. Because of this demonstrated sensitivity, selectivity, and versatility, heterodyned noncollinear XUV wave-mixing techniques will prove invaluable in disentangling electronic dynamics in complex chemical systems.

## Acknowledgements

This work was supported by the Director, Office of Science, Office of Basic Energy Sciences through the Atomic, Molecular, and Optical Sciences Program of the Division of Chemical Sciences, Geosciences, and Biosciences of the US Department of Energy at LBNL under contract no. DE-AC02-05CH11231. A.P.F. acknowledges funding from the National Science Foundation Graduate Research Fellowship Program.

## Tables

| Dark State | LIS – (eV) | LIS + (eV) |
|---|---|---|
| ***1s2s* (20.6 eV)** | 19.1 | 22.1* |
| ***1s3s* (22.9 eV)** | 21.4* | 24.4* |
| ***1s3d* (23.1 eV)** | 21.6 | 24.6* |
| ***1s4s* (23.7 eV)** | 22.2* | 25.2* |
| ***1s5s* (24.0 eV)** | 22.5 | 25.5* |
| ***1s6s* (24.2 eV)** | 22.7 | 25.7* |

**Table 1: Calculated XUV photon energies of key LISs assuming a NIR photon energy of 1.5 eV.** Note that these values incorporate neither the NIR pulse bandwidth nor AC Stark energy shifting of the dark states. LISs observed in this experiment are indicated by an asterisk.

## Figures

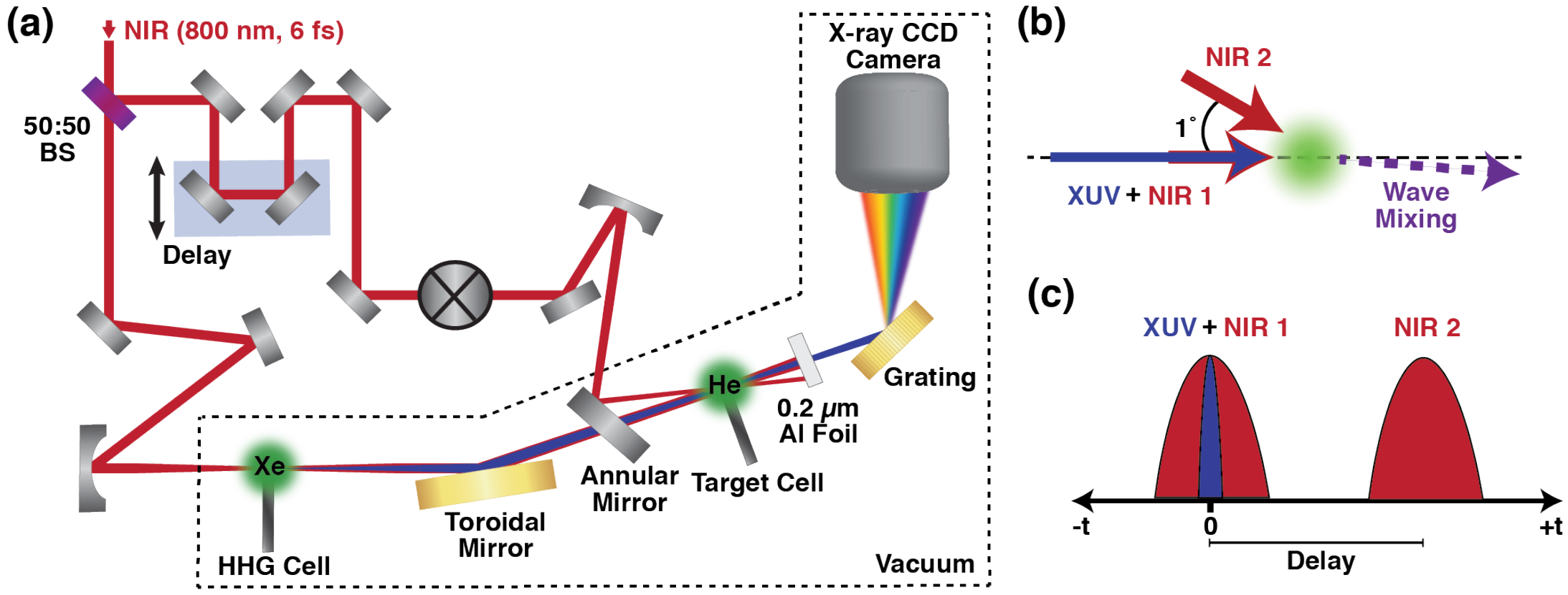

**Figure 1: Experimental geometry and pulse sequence.** (a) The experimental apparatus after spectral broadening and pulse compression used for noncollinear wave-mixing measurements with attosecond pulses. (b) A noncollinear NIR pulse crossing the axis defined by the propagation of the XUV pulse with an angle of 1° leads to wave-mixing emission at non-zero divergence angles. (c) The XUV and collinear NIR pulse are time-coincident and induce a polarization probed by the noncollinear NIR pulse.

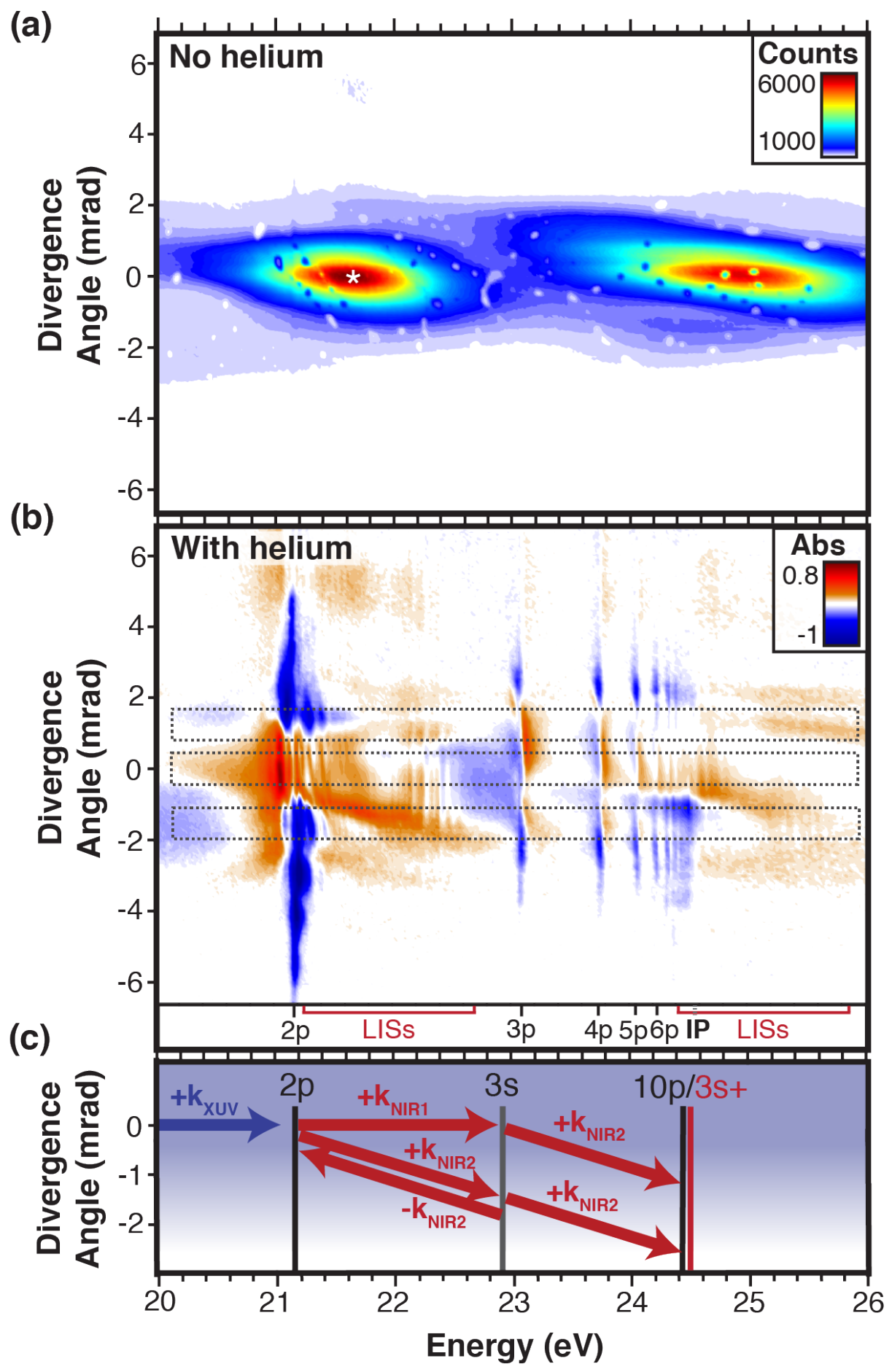

**Figure 2: Noncollinear XUV wave-mixing in a self-heterodyned detection scheme.** (a) A CCD camera image taken of the harmonics at overlap of all three pulses without any gas in the target cell. Note that the harmonics span an angular range of over 4 mrad. The isolated regions with little to no harmonic flux within the harmonic structure arise due to dead pixels and/or debris on the detector. The white asterisk indicates the point of peak XUV flux designated as a divergence angle of 0 mrad. (b) A CCD camera image at temporal overlap with helium gas in the target cell. The dotted boxes indicate regions of the image integrated in the angular dimension for time-domain analysis. Key bright states are assigned below the image. LIS energies are indicated by red brackets. (c) Representative wave-mixing pathways involving one or two noncollinear photons. The shaded blue gradient indicates the angular spread of the harmonics.

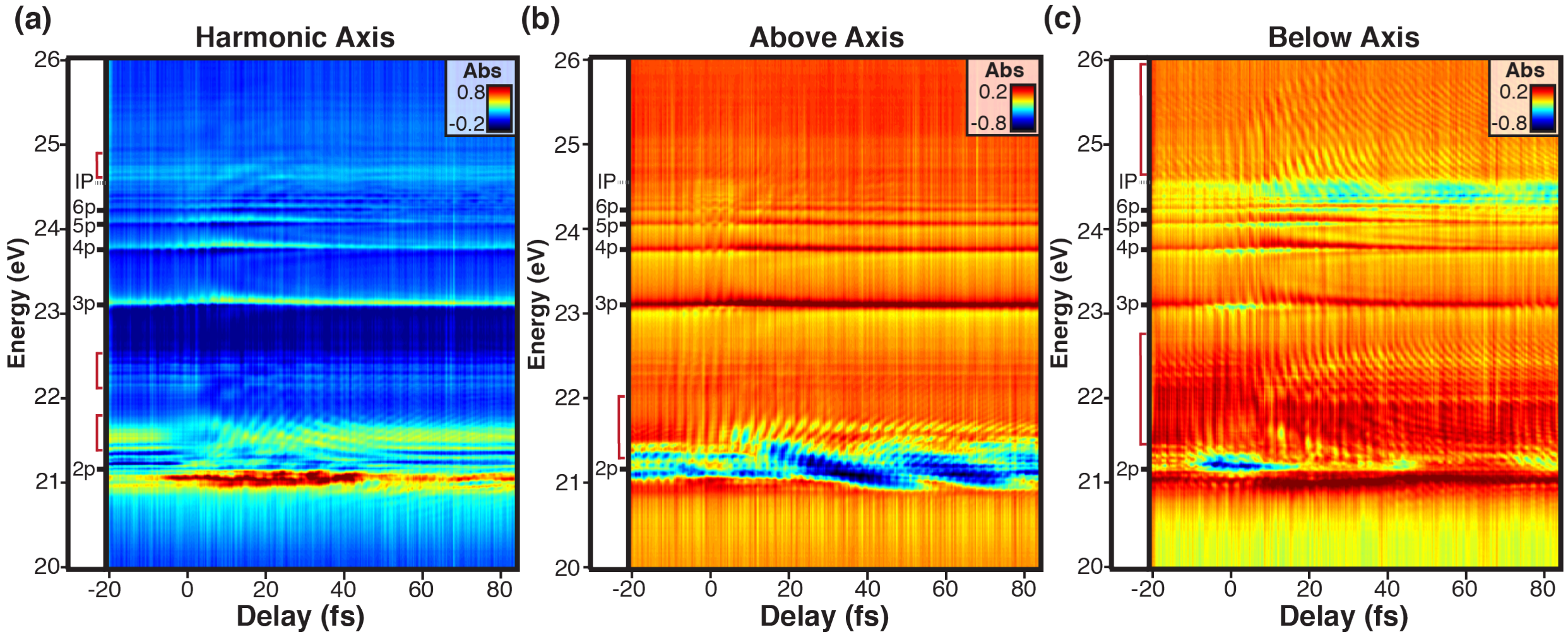

**Figure 3: Transient spectra of four-wave mixing signals emitting from the *1snp* Rydberg states in helium.** (a) Transient spectrum obtained by integrating over the 0.65 mrad region centered on the XUV harmonic axis at a divergence angle of 0 mrad indicated in Fig. 2b. (b) Wave-mixing spectrum obtained by integrating over the region below the harmonic axis indicated in Fig. 2b. (c) Wave-mixing spectrum obtained by integrating over the indicated region above the harmonic axis in Fig. 2b. Key bright states are assigned to the left of each spectra. Red brackets indicate features that can be attributed to LISs.

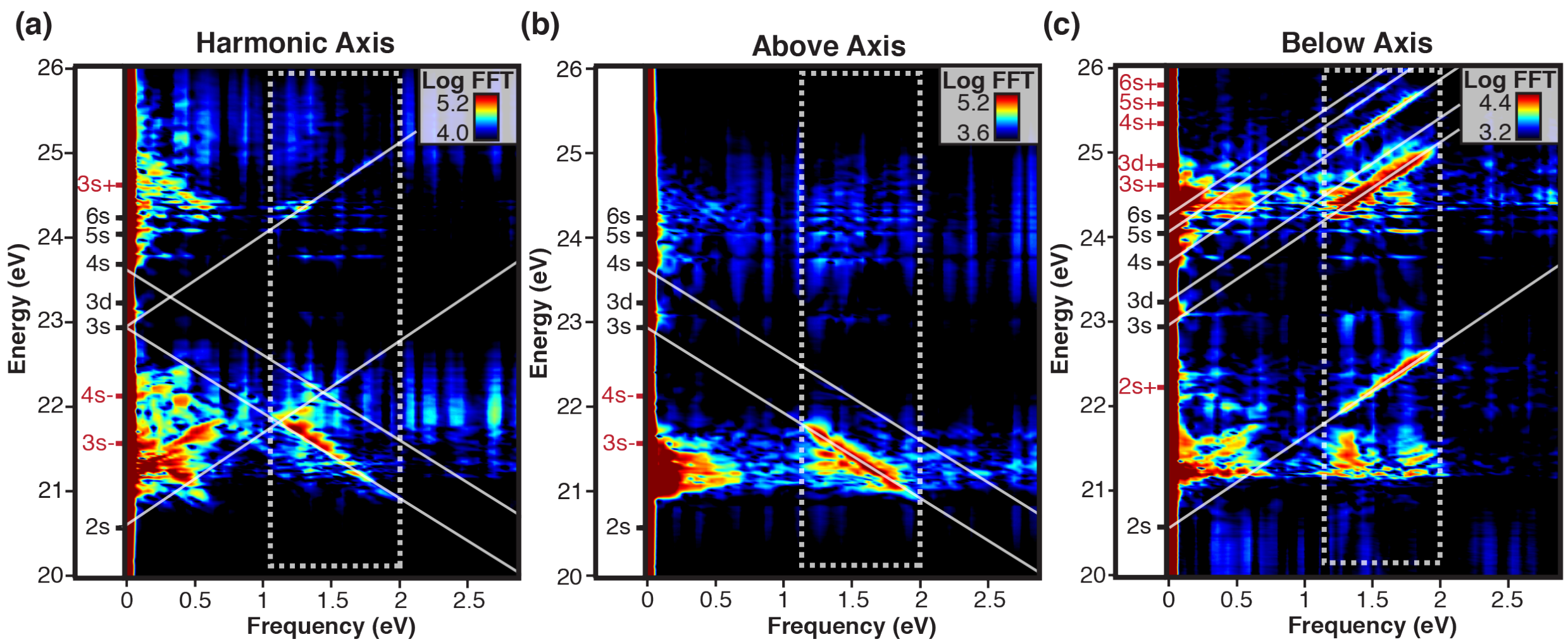

**Figure 4: Fourier analysis of transient wave-mixing spectra.** (a) Fourier transform along the delay axis of the transient absorption spectrum taken on the XUV harmonic axis at a divergence angle of 0 mrad. (b) Fourier transform of the transient wave-mixing spectrum taken below the harmonic axis. (c) Fourier transform of the transient wave-mixing spectrum taken above the harmonic axis. Dotted boxes define the region corresponding to the cycle, ~2.6 fs oscillations. White lines trace the lines of unity slope discussed in the text. Key dark and light induced states are assigned to the left of each spectra.

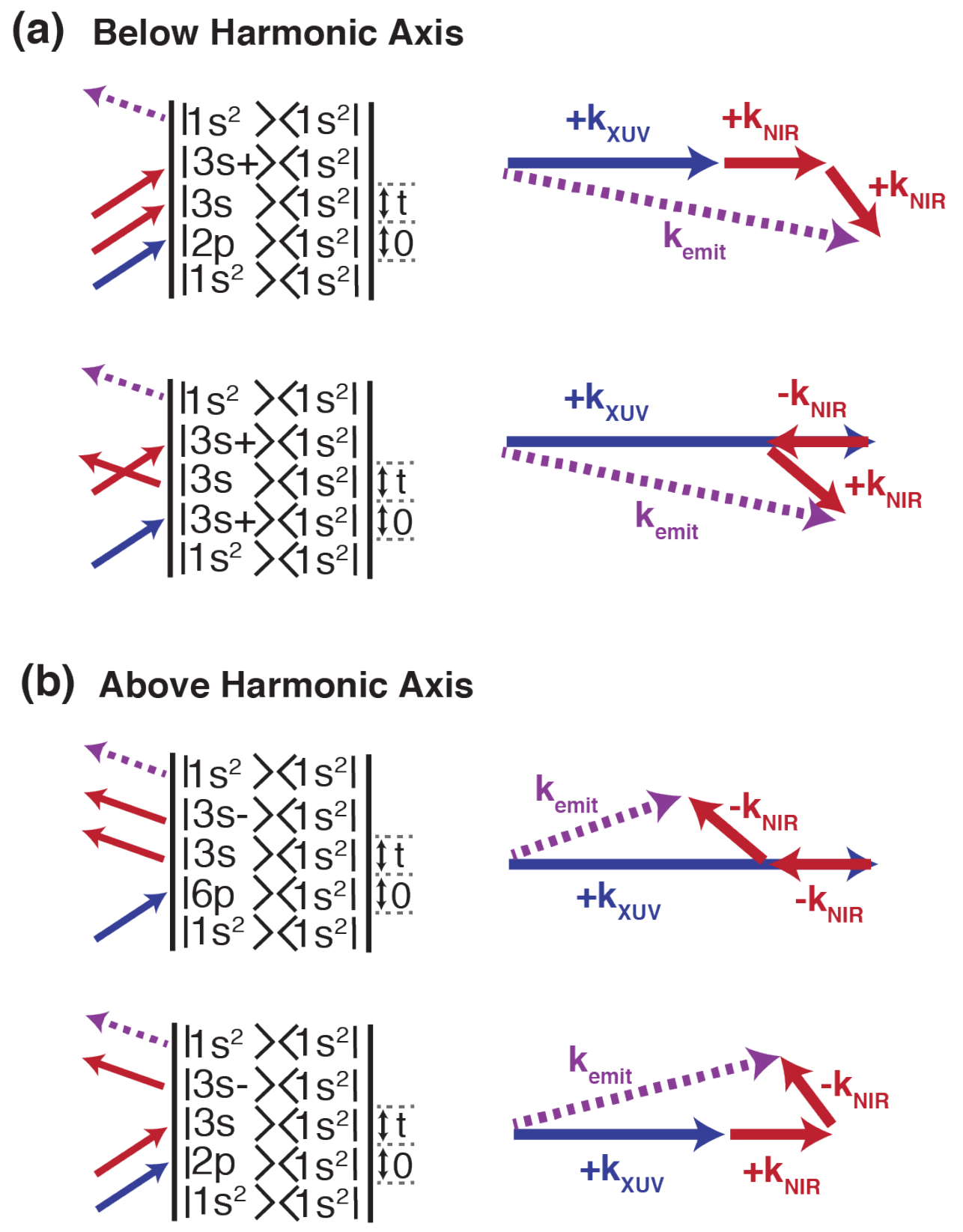

**Figure 5: Double-sided Feynman and wavevector phase-matching diagrams depicting representative wave-mixing pathways exhibiting time-dependence past overlap.** (a) Wave-mixing pathways in which the last interaction consists of the absorption of a noncollinear NIR photon emit below axis. (b) Wave-mixing pathways in which the final interaction can be described as an emission of the noncollinear NIR photon emit above axis. Wave-mixing signals can emit from either LISs or bright states.